\documentclass{egpubl}

\usepackage[T1]{fontenc}
\usepackage{dfadobe}  

\BibtexOrBiblatex
\electronicVersion
\PrintedOrElectronic
\ifpdf \usepackage[pdftex]{graphicx} \pdfcompresslevel=9
\else \usepackage[dvips]{graphicx} \fi

\usepackage{egweblnk} 
\usepackage{amsmath} 
\usepackage{amssymb}
\usepackage[english]{babel}
\usepackage{hyperref}
\hypersetup{  
    urlcolor=blue,
    }
\DeclareMathOperator*{\argmin}{argmin}   

\title[Interpretable Disentangled Parametrization of Measured BRDF with $\beta$-VAE]%
      {Interpretable Disentangled Parametrization of Measured BRDF with $\beta$-VAE}

\author[A. Benamira, S. Shah, S. Pattanaik]
{\parbox{\textwidth}{\centering Alexis Benamira,  Sachin Shah,  Sumanta Pattanaik
        }
        \\
{\parbox{\textwidth}{\centering University of Central Florida
       }
}
}

\begin{document}

 \teaser{
  \includegraphics[width=\linewidth]{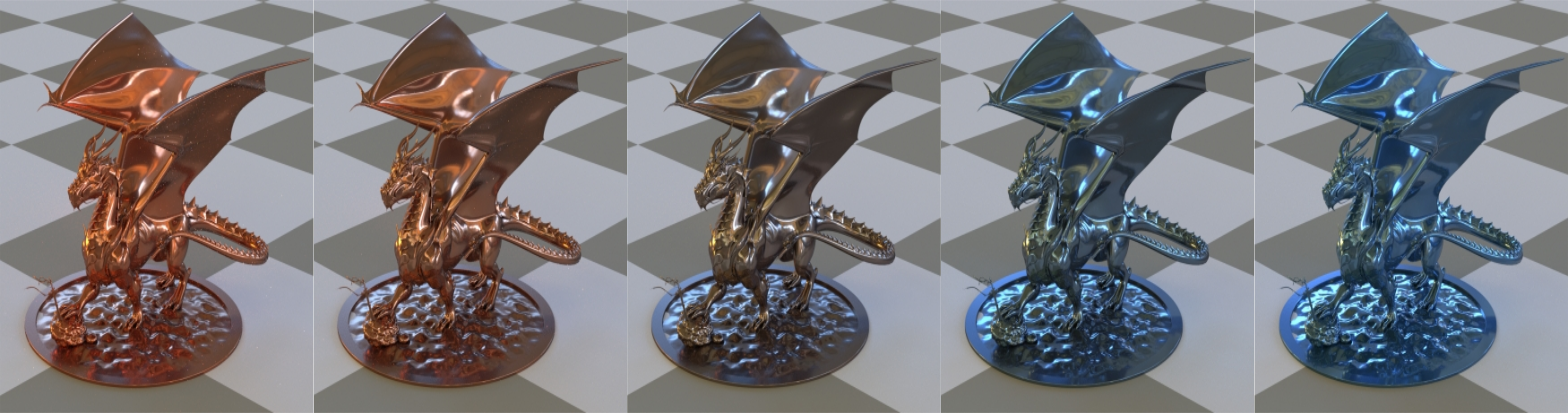}
  \centering
  \caption{
  Latent traversal of parameter $n^{o}8$ of the latent space learned by our Deep Neural Network. All other parameters remain constant. We can visually interpret that this parameter is controlling the specular color of the BRDF, varying from red to blue.}
 \label{fig:teaser}
}

\maketitle
\begin{abstract}
Finding a low dimensional parametric representation of measured BRDF remains challenging. Currently available solutions are either not interpretable, or rely on limited analytical solutions, or require  expensive test subject based investigations. In this work, we strive to establish a parametrization space that affords the data-driven representation variance of measured BRDF models while still offering the artistic control of parametric analytical BRDFs. We present a machine learning approach that generates an interpretable disentangled parameter space. A disentangled representation is one in which each parameter is responsible for a unique generative factor and is insensitive to the ones encoded by the other parameters. To that end, we resort to a $\beta$-Variational AutoEncoder ($\beta$-VAE), a specific architecture of Deep Neural Network (DNN). After training our network, we analyze the parametrization space and interpret the learned generative factors utilizing our visual perception. It should be noted that perceptual analysis is called upon downstream of the system for interpretation purposes compared to most other existing methods where it is used upfront to elaborate the parametrization. In addition to that, we do not need a test subject investigation. 
A novel feature of our interpretable disentangled parametrization is the post-processing capability to incorporate new parameters along with the learned ones, thus expanding the richness of producible appearances. Furthermore, our solution allows more flexible and controllable material editing possibilities than manifold exploration. Finally, we provide a rendering interface, for real-time material editing and interpolation based on the presented new parametrization system.


\end{abstract}  
\section{Introduction}

Natural material appearance is rich and varies over a large set of dimensions. This wide range of possible variations makes material appearance a hard function to represent and developing a corresponding parameter space is just as challenging. A function that is commonly used to represent material appearance is the Bidirectional Reflection Distribution Function (BRDF) and it falls into two main categories: analytical or measured. Analytical BRDFs have been successful at describing a variety of material appearances and several works have developed interpretable parametrizations \cite{burley}, \cite{karis2013real}, \cite{1573631}. Those parametrizations are generally created to be artist-friendly with a representation focus on either a large set of materials or some detailed appearance effect. Unfortunately, analytical models are restricted in the range of appearances they can represent, thus triggering the idea of using measured BRDF. Indeed, with the increasing efficiency and accuracy of BRDF measurements as well as the decreasing cost of storage, datasets keep expanding and data-based reflection models have attracted a lot of attention \cite{Matusik:2003}, \cite{UTIA}, \cite{Dupuy2018Adaptive}. 
However, in contrast to analytical models, parametrization of measured BRDF models remains challenging. Pursuing this issue, three principal lines of investigation have emerged in recent years.

The first line of work propose to accommodate an analytical BRDF model to the measured ones \cite{ravi},\cite{paca},\cite{bagher:hal-00702304}. 
Although quite easy to parameterize, they suffer from two drawbacks. Either the model is too simple to represent the richness of the measured data, or the model is complex and can encounter instability in the fitting procedure.
The second line of work relies on user-based studies to design an interpretable parametrization space \cite{gloss},\cite{translucency},\cite{intuitive}. 
They provide intuitive parametrization space but the test-subject study can be unwieldy to conduct.
The last line of work, that includes the present investigation, makes use of machine learning methodologies. So far, while the offered solutions manage to fit the measured data very well, they do not produce interpretable parametrization solutions \cite{Matusik:2003}, \cite{manifold_para}. 
Close to our work and in an effort to ameliorate BRDF compression, two recent papers apply deep learning methods for faithful compression-decompression of the measured data  \cite{neural_process}, \cite{DeepBRDF}. Using  previously established parametrizations (\cite{intuitive} and \cite{article}), they confer some editing possibility to their work. However, those methods, for being effective compression tools, do not offer the capability of creating an interpretable parametrization space.


In this work, we leverage a Deep Neural Network (DNN) to learn a new interpretable parameter space for measured materials. We rely on a $\beta$-Variational AutoEncoder ($\beta$-VAE), a specific network architecture which enforces the creation of a disentangled latent space. The latent space is the space in which the compressed data lives. Disentangled means that each variable of the latent space is 
responsible for representing a specific generative factor and is insensitive to the factors represented by the other variables. Instead of using perceptual analysis through a cumbersome user study to define an interpretable control space, our set of variables is generated by the DNN through the training process. Then, in a second step, we explore the newly established parametrization space and we exert our visual perception to interpret the generative factors learned by the network.

Expanding on an unsupervised DNN training, our method enjoys the precise fitting and appearance richness of machine learning approaches while offering the interpretability and control of analytical models without requiring crowd-sourced experiments.

The created interpretable parametrization space yields multiple editing applications. First, we can adjoin new parameters to the $\beta$-VAE learned set. The resulting refined set enlarges 
the range of possible appearance representations of our neural network. In addition, it enables the controllable post-tuning improvement of materials directly in image space to better match the initial measured BRDF. Indeed, our neural network output can be rendered and easily improved by changing a few values of our set of interpretable parameters. Because lighting and shape substantially influence our perception of appearance \cite{shape_illu}, being able to improve the quality of the reconstructed BRDF directly in the conditions it will be rendered comes as an important feature. Last, our interpretable parameter space allows more flexible interpolation between materials than random manifold exploring.

To sum up our contributions:

We propose a $\beta$-VAE based approach to establish a new interpretable editing parametrization space for measured BRDFs models. Instead of resorting to a user study and collecting empirical observations, we train a DNN in an unsupervised way to efficiently generate a disentangled latent space of generative factors that are highly interpretable for humans.

We apply our newly found parametrization space for various tasks: reconstruction improvement, material editing with flexible interpolation as well as random manifold exploring and measured BRDF compression.

We provide two interfaces for material editing based on BRDFExplorer \cite{burley}, one, real-time, using our newly found paramerization space and the other for random manifold exploring. We can experience firsthand the higher control flexibility of our parametrization over the manifold exploration.

To the best of our knowledge, our work is the first report of a deep learning approach applied to create an interpretable parameter space for measured BRDF models.

The organization of the rest of the paper is as follows. First, we offer an overview of the state of the art in measured BRDF parametrization and other relevant current research. Then, we describe our method followed by our results before concluding.

\section{Related Work}

\subsection{Measured BRDF Parametrization}

\textit{Measured BRDF datasets}. Because analytical modeling of material appearance is complex, measured BRDF models have always attracted a lot of attention. Measurement techniques have become more accurate and simpler \cite{10.1145/2816795.2818085}. Nowadays, BRDF measurements can merely be accomplished using a smart phone camera and a flash \cite{deschaintre2018single}. Various datasets of BRDF measurements are available \cite{UTIA}, \cite{Dupuy2018Adaptive}. In this work, we decided to use the MERL dataset \cite{Matusik:2003} as it offers a wide set of material measurement data. We will discuss in more details the specifics of the MERL BRDF data in Section~\ref{subsect:data_prepro}. 

\textit{Analytical model fitting}. As touched upon earlier, finding a parametrization for measured BRDF models has been achieved using three main approaches. The first by fitting an analytical model to the measurements. The goal of those efforts are twofold, first to improve the accuracy of the analytical model by comparing them to measured data \cite{paca}, \cite{brdfanalysis}, \cite{analytic1}, \cite{bagher:hal-00702304}, \cite{10.1145/2907941}. Second is the possibility to edit measured BRDF by editing the fitted analytical model \cite{ravi}, \cite{10.5555/2858834.2858838}, \cite{Lawrence:2006:IST}. The primary challenge of this approach is to develop an analytical model complex enough to represent the richness of the measured data, yet simple enough so that the fitting process remains stable.

\textit{User study investigations}. A second method to parameterize measured BRDF calls for user studies and perceptual data gathering. This approach has the inherent advantage of offering a parametrization that is constitutively interpretable. It has been reported for a subset of appearance characteristics: by \cite{gloss} and \cite{10.1145/344779.344812} on glossy reflections or \cite{translucency} for translucency. It has also been studied for particular material types, like in \cite{perc_metal} for metals. In a pioneering work, \cite{intuitive} offered an editing parametrization for all measured BRDF of the MERL dataset for all appearance characteristics. They relied on a Principal Component Analysis (PCA) dimension reduction of the initial data. Later, \cite{manifold_para} and \cite{DeepBRDF} showed that dimension reduction of the measured BRDF with PCA requires retaining a large set of dimensions for good reconstruction of the material appearance, thus, making it impractical to use for parametrization. More recently, \cite{shape_illu} offers an extensive user study to analyze the influence of shape and illumination on the perception of material appearance.

\textit{Machine learning}. Machine learning approaches have been used to find manifold representations of measured BRDF data. In the original paper introducing the MERL database, a non linear dimension reduction techniques named charting \cite{charting} is already being used \cite{Matusik:2003}. By analyzing where groups of similar materials are located on the manifold, they can find paths between those groups to edit the material appearance. The approach to find the manifold as well as the interpolation between material has been refined in \cite{manifold_para} by relying on a Gaussian Process. However the editing is performed with pseudo-random manifold exploration. It is pseudo-random because the material clustering obtained on the manifold allows general control over the interpolation editing but does not allow precise control over specific appearance characteristics. In two recent works, deep learning approaches oriented toward compressing the input measurements have been presented \cite{DeepBRDF}, \cite{neural_process}. Both offer some editing possibilities by using the parametrization offered in \cite{article} and \cite{intuitive}, but their approach does not allow the creation of an interpretable parametrization over the measured BRDF models. Current deep learning approaches lack interpretability. 

In this work, we endeavor to alleviate the deep learning black box effect by finding an interpretable editing parametrization space for measured BRDF model. The newly generated editing parametrization is obtained without any user study, merely through the training of a DNN. Visual perception is applied in a second step to interpret the parametrization obtained by the network. Our goal is to find a parametrization space that offers at the same time, the artistic control of the analytical BRDF models while and the ability to represent the variance of measured BRDF models.


\subsection{Generative Deep Neural Networks for materials}

Deep Learning is a fast growing area of research. It has been very successfully applied to graphics in a wide variety of areas and specially rendering. \cite{neural_rendering} and \cite{tewari2021advances} present a thorough review of the latest progress in neural rendering. In recent years, DNN solutions have been useful in generating Spatially-Varying-BRDF from pictures of a surface \cite{deschaintre2018single}, \cite{10.2312:sr.20201136}, \cite{10.1145/3450626.3459854}. To solve this problem, several works have focused their attention onto the latent space in order to improve the performances of their approach \cite{latent_brdf1}, \cite{latent_brdf2}. However so far, none has discussed the possibility of finding a disentangled latent space. In our work, we also pay careful attention to the latent space, forcing its disentanglement in order to create an interpretable parametrization of the measured BRDF. Other examples of applications include \cite{zsolnaifeher18gms} where a deep learning approach is used to generate a fixed image with different material appearances, but the control over the generative process is done through a random manifold exploration and therefore lacks interpretability. In a very recent work, \cite{delanoy2022generativematerials} improves on the latter work by offering interpretable editing of materials on images with a DNN. They rely on a crowd-sourced experiment for perceptual data gathering.

\subsection{Disentangled Deep Neural Networks}
One major drawback of Deep Neural Networks is their behavioral opacity. To offer some interpretability on the learning process a growing amount of AI research tries to develop models offering a disentangled and compressed representation of the input data \cite{hu2021architecture}, \cite{ren2021interpreting}, \cite{NEURIPS2020_6fe43269}. Indeed, disentangled representation of data usually have good human interpretability. Simply put, each variable encodes a generative factor that is perceptually meaningful. According to the definition of \cite{10.5555/3326943.3326961}, a disentangled representation should embrace three elements, modularity, compactness and explicitness. Modularity depicts the fact that a generative factor is only encoded by a single latent variable and not shared by the others. Compactness signifies that no two variables represent the same generative factor. Explicitness represents the fact that all the generative factors are captured. Several different architectures have been described to find such a disentangled representation \cite{betavae}, \cite{NIPS2016_7c9d0b1f}, \cite{kim2018disentangling}. We picked the $\beta$-VAE architecture \cite{betavae} for this work as it is the most studied one.

\section{Background}

In this section we offer some background on autoencoders, VAE and $\beta$-VAE architectures of DNN. \cite{DBLP:journals/corr/abs-1804-03599}, \cite{hinton2006reducing}, \cite{weng2018VAE} and \cite{DBLP:journals/ftml/KingmaW19} thoroughly cover the latest developments on the subject.

\subsection{Autoencoder}

An autoencoder is a DNN architecture that is composed of two main blocks. First, an encoder $g_{\phi}$ that takes an input $x$ and compresses it into a latent vector of smaller dimension $z$. The second block is called a decoder noted $f_{\theta}$. It take the latent vector $z$ as input and generates an output $\tilde{x}$. $\phi$ and $\theta$ are the parameters of the encoder and decoder, respectively. 

\begin{equation}
	g_{\phi}(x) = z \quad \textrm{and} \quad f_{\theta}(z) = \tilde{x}
\end{equation}

The goal of training an autoencoder is to find $\hat\phi$ and $\hat\theta$ such that the output $\tilde{x}$ is as close as possible to $x$ along the norm $|| . ||$ defined by the user. 
\begin{equation}
	\hat\phi, \hat\theta = \argmin_{\phi,\theta}(|| x - f_{\theta}(g_{\phi}(x)) ||)
\end{equation}

The latent vector z can then be used as a compressed representation of the input x. This architecture has been successfully applied to various tasks such as denoising or classification \cite{10.1145/1390156.1390294}, \cite{lin2013spectral} and has been used in \cite{DeepBRDF} for BRDF compression. However, this architecture has limitations for generative applications. Indeed, because of the lack of constraints on the latent space, the continuity between the latent representations of the different inputs cannot be guaranteed, and the interpretability of each encoded dimension is not possible. To address the continuity issue, Kingma and Welling introduced the Variational Autoencoder (VAE) architecture \cite{Kingma2014}.

\subsection{Variational Autoencoder (VAE)}
With a VAE, instead of encoding the input into a point in the latent space, it is mapped to a distribution. This distribution is named prior, noted $p_{\alpha}(z)$ and depends on some variables $\alpha$. For example, with a gaussian distribution, $\alpha$ is the mean and standard deviation. This prior distribution is chosen by the user. 
In our architecture, it is the encoder that generates the latent space. So the goal of the encoder, now a probabilistic encoder noted $g_{\phi}$, is to generate a distribution that matches the prior distribution. The latent vectors will then be sampled from the distribution generated by the encoder. We want to find the parameters $\phi$ such that for every input of the encoder $x$ the output $z$ follows the prior distribution: $g_{\phi}(z|x) \equiv p_{\alpha}(z)$. The decoder, now a probabilistic decoder noted $f_{\theta}$ takes the latent vector and tries to reconstruct the input as closely as possible. We want to find the parameters $\theta$ which maximize $f_{\theta}(x|z)$, the likelihood of reconstructing $x$ given $z$ which is sampled from the probabilistic encoder distribution.

In other words, in one hand we want the distribution given by the probabilistic encoder to be as close as possible to the prior distribution that is set by the user. On the other hand we want the decoder to maximize the likelihood distribution of reconstructing back any input $x$ from the corresponding latent sample $z$ where $z$ is obtained from the probabilistic encoder. The first element enforces a latent distribution, the second part enforces an identical reconstruction of the input. This can be formalized as follows \cite{Kingma2014}:

\begin{equation}
	\hat\phi, \hat\theta = \argmin_{\phi,\theta}(D_{KL}(g_{\phi}(z|x)||p_{\alpha}(z)) - \mathbb{E}_{z \sim g_{\phi}(z|x)}\log(f_{\theta}(x|z))
)
\label{eq:lossvae}
\end{equation}

The first term enforces the probabilistic encoder to generate the prior distribution by finding $\phi$ which minimises the Kullback–Leibler Divergence, noted $D_{KL}$, between $p_{\alpha}(z)$ and $g_{\phi}(z|x)$. The second term enforces the reconstruction, that allows us to find $\theta$ that maximizes the log-likelihood of generating $x$ from $z$ where $z$ is sampled from $g_{\phi}(z|x)$.

\subsection{$\beta$-Variational Autoencoder ($\beta$-VAE)}
The VAE architecture and loss function ensure that the latent space is continuous by learning a latent distribution while trying to reconstruct the input as faithfully as possible. The $\beta$-VAE modifies the loss function defined in Equation~\ref{eq:lossvae} slightly through the inclusion of a multiplier term $\beta$ to the KL divergence part.

\begin{equation}
\begin{gathered}
\mathcal{L}_{\beta VAE} =
\beta*D_{KL}(q_{\phi}(z|x)||p_{\theta}(z)) - \mathbb{E}_{z \sim q_{\phi}(z|x)}\log(p_{\theta}(x|z))
\end{gathered}
\end{equation}

$\beta=1$ yields the original VAE loss function. Setting $\beta>1$ puts more emphasis on the latent distribution for a better encoding. The most efficient encoding for conditionally independent factors is orthogonal. So emphasizing on the latent distribution encourages the generative factors to be represented in a disentangled fashion.

In the present work, we create a new DNN following a $\beta$-VAE architecture to learn a disentangled latent space over measured BDRF. In the next section, we will discuss the details of our architecture and training as well as the input pre-processing.

\section{Method}

\subsection{Data Pre-processing}\label{subsect:data_prepro}

The MERL dataset \cite{Matusik:2003} is composed of the measured BRDF of 100 isotropic materials. Measurements are densely sampled using Rusinkiewicz half-difference angles $\left(\theta_h, \theta_d, \phi_d\right)$ \cite{Rusinkiewicz98}. The raw data is stored in a 3-dimensional table shaped $\left(90, 90, 180\right)$ with high dynamic range RGB values for each entry. These high magnitude inputs can pose challenges for a DNN, so the following normalization scheme is adopted:

\begin{equation}
\hat\rho = \frac{\log\left(\rho * S + \varepsilon\right) - \log\left(\varepsilon\right)}{\log\left(1 + \varepsilon\right) - \log\left(\varepsilon\right)}
\end{equation}

where $\rho$ is a single BRDF measurement, $\varepsilon = 0.01$ is a small constant and $S$ is a scalar value used to render the MERL BRDF. $S$ is $\frac{1.0}{1500}$ for the red channel, $\frac{1.5}{1500}$ for the green channel, and $\frac{1.66}{1500}$ for the blue channel. Additionally, regions of the BRDF where the viewing direction or incident direction are below the horizon are set to $0$ indicating an invalid entry \cite{burley}. The BRDF is treated as $180$ RGB image slices shaped $90 \times 90$. 
To reduce the amount of data at the entry of our network, we discard 8/9 of the original slices at regular interval. This ratio is chosen following \cite{DeepBRDF}. We now have 21 RGB images slices, producing $63 \times 90 \times 90$ input size. The reconstruction of a full BRDF is obtained through linear interpolation between image slices. 

\subsection{Network Architecture}

\begin{figure}[h]
      \centering
      \includegraphics[width=.9\linewidth]{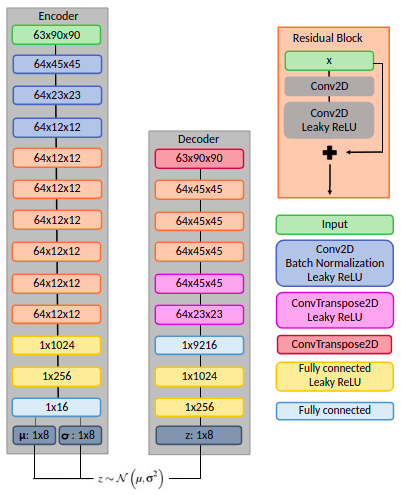}
      \caption{\label{fig:network}
               Our $\beta$-VAE architecture.}
\end{figure}

We expand on the autoencoder introduced by \cite{DeepBRDF} to create a new $\beta$-VAE to learn a disentangled latent space. The proposed $\beta$-VAE architecture is illustrated in Figure~\ref{fig:network}. We choose a normal distribution $\mathcal{N}\left(0, 1\right)$ as a prior. The encoder $g_{\phi}$ generates a mean and a variance $\mu_{x}, \sigma_{x}$ for each input $x$. A single point $z$ is sampled from the normal distribution $\mathcal{N}\left(\mu_x, \sigma_x^2\right)$.

\begin{align}
      g_{\phi}(x) &= [\mu_{x}, \sigma_{x}^2]\\
      z &\sim \mathcal{N}\left(\mu_{x}, \sigma_{x}^2\right) \label{equ:repram}\\
      f_{\theta}(z) &= \tilde x \label{eq:recon}
\end{align}


Our proposed network is composed of several 2D convolutional layers. They all have a kernel size of $3 \times 3$, stride length of 2, and padding size of 1 except the last one of the decoder that has a kernel size of $4 \times 4$. Batch normalization and leaky ReLU activation are performed after each layer as indicated Figure~\ref{fig:network}. Nine residual blocks \cite{He_2016_CVPR} with two 2D convolution layers (kernel size of $3 \times 3$, stride length of 1, and padding size of 1) with leaky ReLU activation are added. Then three fully connected layers are used to extract the mean and a standard deviation of the latent space distribution. The latent vector is chosen of size 8. To decode the latent vector, a point is sampled from the distribution and expanded by three fully connected layers. Before entering the first 2D convolutional layers of the decoder, the vector is reshaped to a $64\times 12\times 12$. Finaly, after three residual blocks, and a last convolutional layer we get the output result in the same sliced BRDF format as the input.

\subsection{Implementation Details}

Our $\beta$-VAE network is trained in an unsupervised way and implemented using PyTorch \cite{NEURIPS2019_9015}. The network is trained with the Adam optimizer \cite{DBLP:journals/corr/KingmaB14}, a learning rate of $3\times10^{-5}$, and batch size of $2$ for $1000$ epochs. Larger batch sizes failed in faithfully reproducing the color of the materials, leaving all of them brown.  Equation~\ref{equ:repram} is replaced with the reparametrization trick \cite{Kingma2014} during training.
\begin{equation}
      z = \mu + \sigma \times \epsilon \quad \textnormal{with} \quad \epsilon \sim \mathcal{N}\left(0, 1 \right)
\end{equation}

The loss function is formulated as
\begin{align}
\mathcal{L}_{recon}^{(i)} &= \lVert mask(\tilde x^{(i)}) - mask(x^{(i)}) \rVert\\
\mathcal{L}_{KL}^{(i)} &= -\frac12 \sum_{j=1}^{M} \left(1 + 
\log{ \left( \sigma_j^{(i)} \right)^2 } - \left(\mu_j^{(i)}\right)^2 - \left( \sigma_j^{(i)} \right)^2  \right)\\
\mathcal{L} &= \frac{1}{N}
\sum_{i=1}^{N} (\mathcal{L}_{recon}^{(i)} + \beta_{norm}\mathcal{L}_{KL}^{(i)})
\end{align}
where $M$ is the dimensionality of the latent space, $N$ is the mini-batch size, $x$ is an input BRDF, $\tilde x$ is the reconstruction from Equation~\ref{eq:recon}. $\lVert \cdot \rVert$ denotes the $L_2$ norm. The function $mask$ returns the set of all valid BRDF entries. $\beta_{norm} = \beta \frac{M}{N}$ where $\beta = 12$, $M = 8$ is the latent space size and $N = 63 \times 90 \times 90$ is the input size.

Concerning the dimension of the latent space, \cite{DeepBRDF} have shown that $M=10$ is a cut-off dimension: smaller latent space deteriorate the reconstruction greatly, while larger dimensions produce limited gains. We trained a DNN with $M=8$ and $M=16$. When $M=16$, the explicitness of the network did not improve and we faced the same poor reconstruction with green materials. Furthermore, several parameters were not interpretable or did not seem to have noticeable control over the appearance and multiple parameters seemed to control the same appearance characteristic. Hence, we decided to use $M=8$ to improve the disentanglement by further constraining the latent representation.

It can be noted here that by construction, the $\beta$-VAE architecture enables the compression/decompression of the input data. In our case, the $63 \times 90 \times 90$ input is compressed into an 8 elements vector. The storage of each material of the MERL dataset can be replace by the storage of each associated pair $\left(\mu, \sigma\right)$ and the decoder of our $\beta$-VAE. After training, the size of our model used to run our interfaces is $78.9$ MB, the size of the latent vectors are negligible. For comparison, the size of one of the MERL measurement is $35$ MB. 

\begin{figure*}[tbph]
  \centering
  \mbox{} \hfill
   
  \includegraphics[width=0.98\linewidth]{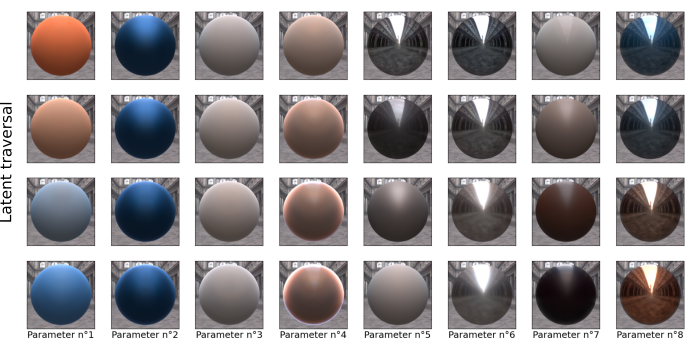}
  \caption{\label{fig:latent_traversal}Latent traversal of each variable of the latent space. We notice that each variable controls a distinct generative parameter. The names associated to each parameter are listed in Table~\ref{tab:latent_traversal_names}. For metallic materials unchanged values are set to the values obtained for the aluminium BRDF measurement. For diffuse materials, only Parameter $n^{o}$5 is changed in the aluminium set for illustration purposes. For Parameter $n^{o}$2, we chose the blue-fabric BRDF value for better contrast with Parameter $n^{o}$3.}
\end{figure*}

\subsection{Interfaces}
For practical handling of the disentangled parametrization, we developed an applicative interface based on BRDFExplorer \cite{burley} to which we added Pytorch C++ and CPython. 
As described earlier, for each BRDF measure, two vectors $\mu$ and $\sigma$ are produced by the encoder.
Our new interface loads a pair of vectors $\left(\mu, \sigma\right)$ representing a BRDF and samples a vector $z$ from the normal distribution of parameters $\left(\mu, \sigma\right)$.
We then propagate $z$ through the decoder of our $\beta$-VAE and render the output BRDF. Our interface runs in real-time and enables the user to modify each value of the latent vector $z$, leveraging the full strength of our parametrization. Indeed, each dimension of the latent vector $z$ represents one of our learned interpretable parameter over the measured BRDF. As described and illustrated in the Results Section, each dimension controls its own interpretable generative factor. Our interface includes sliders to modify  the value of each parameter thus enabling controlled editing of the material. A demonstration video is available \href{https://www.veed.io/view/1d9fd572-2890-4360-9f36-0f4b1bebf2ff}{here}\footnote{Videos are fully anonymized. Link to the code for the interface has been removed to preserved anonymity}.



In addition, we put together a second interface that relies on manifold exploration for editing $z$. This interface is created for comparison sake with the previous related techniques \cite{Matusik:2003}, \cite{manifold_para}, \cite{neural_process}. The latent vector $z$ is mapped to a 2D embedding using Uniform Manifold Approximation and Projection (UMAP) \cite{umap-software}. UMAP is particularly useful because the algorithm preserves the global data structure and is invertible. A 2D point can be dragged over the manifold and mapped back to an 8-dimensions latent vector that is then passed through the decoder before rendering. We observe that this interface has limited editing control over generated material appearance as illustrated in another demonstration video available \href{https://www.veed.io/view/4bb54f48-b0ea-4989-8328-822ab179a3bc}{here}.

\section{Results}

In this section, we illustrate the applicability of our approach and detail the outcomes. First and foremost, we show that we have successfully learned a highly interpretable editing parametrization space from the measured MERL BRDF dataset. Second, we illustrate that our interpretable disentangled parametrization affords the introduction of 
new generative parameters to the $\beta$-VAE learned set to expand the richness of possible material produced. Third, we demonstrate that our parametrization space allows controllable and flexible interpolation between materials as well as new material creation. We used Mitsuba renderer \cite{Mitsuba} to render our images and we obtained some of the scenes used to illustrate our work from \cite{resources16}. 

\subsection{Interpretable Editing Parametrization}

Our goal is to learn an interpretable parametrization for measured BRDF model. To evaluate our parametrization, we perform a latent traversal of each of the 8 dimensions of our latent space. In other words, we modify the value of one of the latent parameter while leaving the others unchanged and render an image for each new value. In a second step, we exert our visual perception to express what we believe is the generative parameter being controlled by the latent variable.
We exercise our perception only as a mean of interpretation of the parameters learned by our DNN. The learning process is unsupervised and does not require any human intervention. 

The results of the latent traversal are presented Figure~\ref{fig:latent_traversal}. We observe that each parameter controls its own generative factor. By relying on our visual perception and using the same terminology as in \cite{burley}, we associate a name to each individual parameters. The naming association is detailed in Table~\ref{tab:latent_traversal_names}.


\begin{table}[h]
\begin{center}
\begin{tabular}{|c |c |} 
 \hline
 Parameter $n^{o}$ & Controlled generative factor \\ [0.5ex] 
 \hline\hline
 1 & Diffuse color from blue to red \\
  \hline
 2 &  Sheen  \\
 \hline
 3 & Subsurface  \\
 \hline
 4 & Clear coat \\  
 \hline
 5 & Specular to Diffuse \\
 \hline
 6 & Haziness  \\
 \hline
 7 & Color lightness  \\
 \hline
 8 & Specular color from red to blue\\  
 \hline
\end{tabular}
\caption{\label{tab:latent_traversal_names} Using the same terminology as in \cite{burley}, we associate a name to the generative factor controlled by each parameter. The association is achieved through visual perception.}
\end{center}
\end{table}

The parameter naming process can be conducted easily and without ambiguity. Thus, we can confirm that the parametrization found is fully interpretable.

\subsection{New Parameters Creation for Expanded Material Appearance Richness}

\begin{figure}[h]
  \centering
  \includegraphics[width=.9\linewidth]{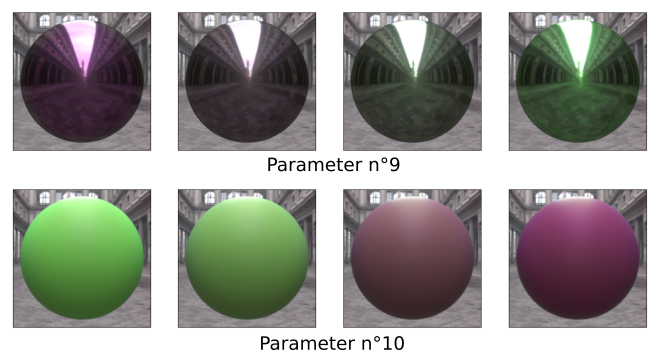}
  
  \caption{\label{fig:human_made_parameter}
           Manual enrichment of DNN-learned parameter set. New manually-created parameters control respectively the green specular and green diffuse color. When the green values are low, only red and blue remain thus creating purple.}
\end{figure}

As detailed in the previous section, we have established an interpretable disentangled parametrization for the measured BRDF of the MERL dataset. The disentangled component is important as we can see that only two parameters control the color of the generated material: parameters $n^{o}1$ and $n^{o}8$, which represent respectively the diffuse and specular color from blue to red. In order to increase the color gamut of our original parametrization, we included two new parameters to the set learned by our DNN. These two new parameters control the diffuse and specular color from green to purple as observed in Figure~\ref{fig:human_made_parameter}.

The creation of these new parameters is achieved as follows. We generate two materials $M1$ and $M2$ from two sets of parameters $V1$ and $V2$ which have the same values for all parameters except for those controlling the color, i.e. parameters $n^{o}1$ and $n^{o}8$. As discussed previously, the two materials follows the RGB format. We replace the Green channel of $M1$ with the Red channel of $M2$, thus creating a new material $M$. Because parameters $n^{o}1$ and $n^{o}8$ of the set $V2$ initially controlled the red diffuse and specular color of $M2$, through this replacement, they now control the green diffuse and specular color of the new material $M$. Next, we add the parameters $n^{o}1$ and $n^{o}8$ of $V2$ to the set $V1$ creating a new set $V$ parametrizing material $M$. 
Those two parameters, now numbered 9 and 10 are responsible for controlling the green color for diffuse and specular reflections as illustrated in Figure~\ref{fig:human_made_parameter}.

\begin{figure}[h]
  \centering
  \includegraphics[width=.95\linewidth]{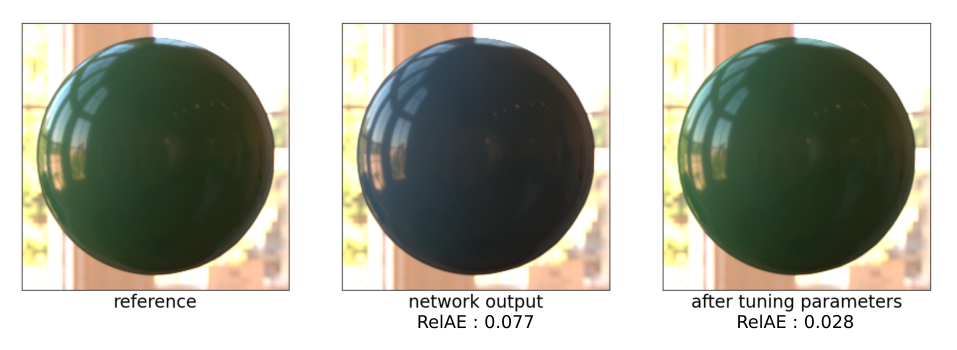}
  \includegraphics[width=.95\linewidth]{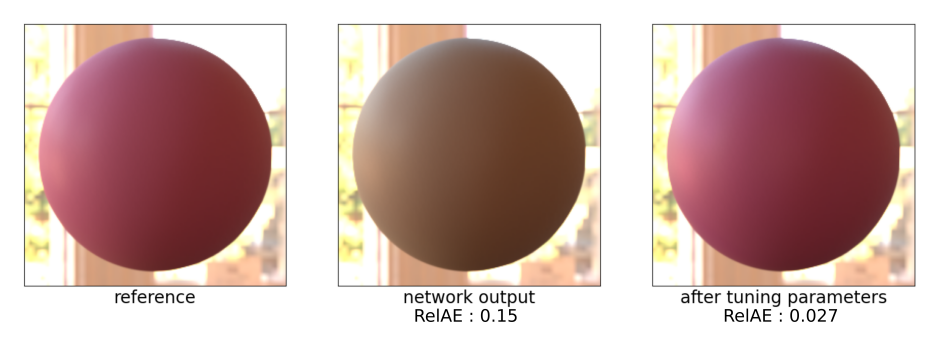}
\includegraphics[width=.95\linewidth]{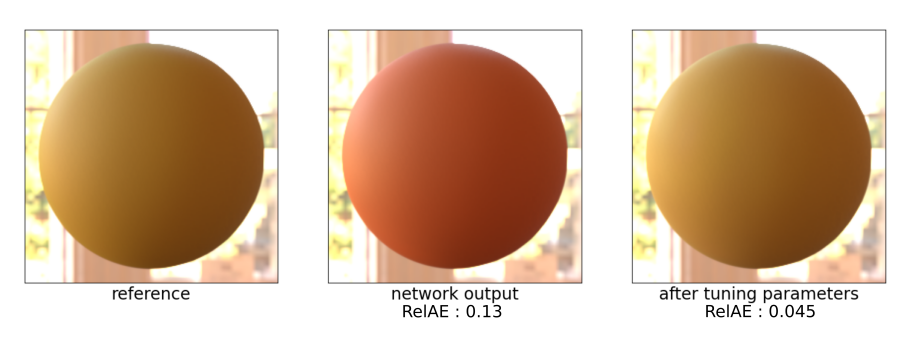}

  \caption{\label{fig:reconstruc_amelio}
           Broadening the range of generated material appearances. With the post-processing manual addition of parameters, we improve the range of possible material appearances generation of our network and enhance the DNN reconstruction of the original materials of the MERL dataset. We use the Relative Absolute Error (RelAE) metric, the lower the better. Top: green-acrylic, Middle: violet-rubber, Bottom: yellow-paint.}
\end{figure}

This post-processing manual addition of parameters is only possible because the initial parametrization learned by our DNN is interpretable and disentangled. Indeed, the disentanglement feature forces the color to be controlled by two parameters only and the interpretability allows us to modify the original purpose of the parameters. The two newly created parameters allow us to broaden the range of generated material appearances. As illustrated Figure~\ref{fig:reconstruc_amelio}, we can now reproduce more faithfully the original material appearances of the MERL dataset by tuning our new set of parameters.

\subsection{Flexible Material Editing}

\begin{figure}[h]
  \centering
  \includegraphics[width=.90\linewidth]{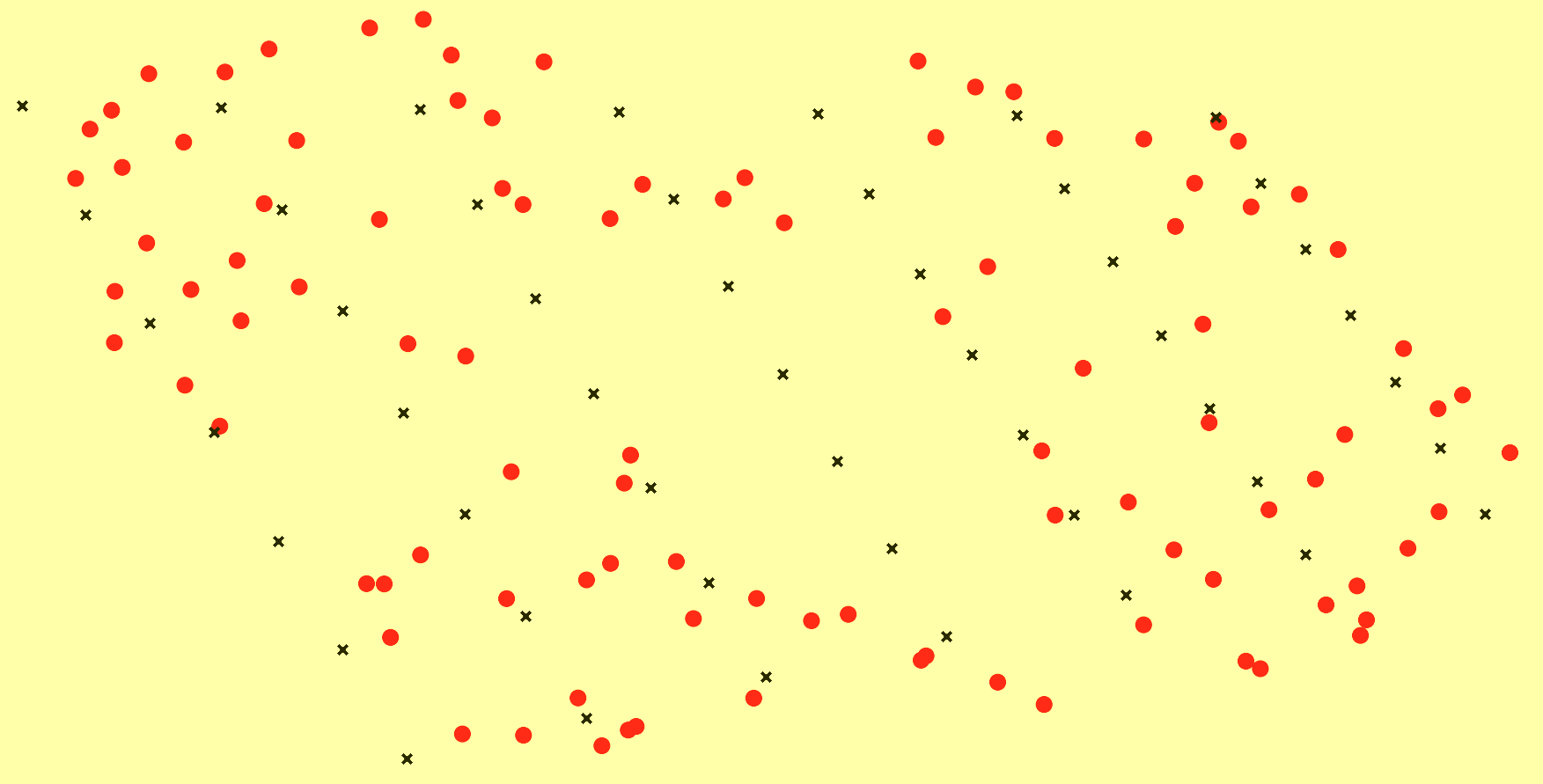}
  \includegraphics[width=.90\linewidth]{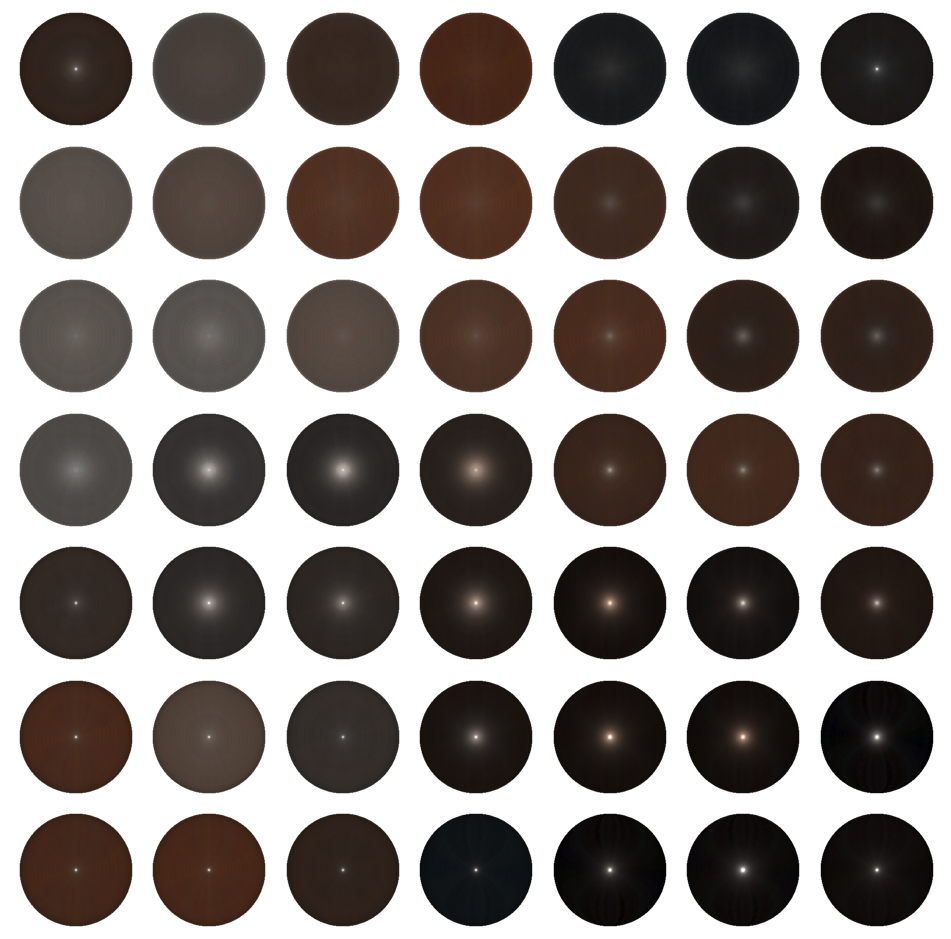}

  \caption{\label{fig:umap}
  Previously available manifold exploration methods are impractical for editing.
           Top: UMAP manifold representation of the latent space. UMAP enables the representation of the latent space on a 2D plane. The latent vectors corresponding to the materials of the MERL dataset are represented by red dots, the $7 \times 7$  sampled points are represented by black crosses.
           Bottom: Rendering of the sampled manifold points. We render the 49 sampled points with a directional light source. We can identify some general tendencies over the locations of similar materials. For example, brighter colors are on the left and darker colors on the right. Exploration of the manifold for material editing can be troublesome to control.}
\end{figure}

Previously available machine learning approaches relied on finding a path on a learned manifold for editing the measured materials. For comparison purposes we did project the latent space learned by our $\beta$-VAE onto a 2D manifold using the UMAP technique \cite{umap-software}. We then sampled $7 \times 7$ points on the found manifold and rendered the materials associated to each point. The results can be visualized in Figure~\ref{fig:umap}. We identify some general appearance areas on the manifold. For example, the upper part is more diffuse and the bottom part is more specular. Navigating between the points allows some editing options. However, it proves quite challenging to interpolate between material that are far apart on the manifold. Indeed, finding a path that interpolates faithfully the appearance between materials that are very different, i.e. far on the manifold, requires discovering a long path on the manifold. Moreover, it appears that on some paths small steps on the manifold can create abrupt changes on the material appearance as illustrated in the companion video demonstrating the UMAP interface (see previous link).

\begin{figure}[h]
  \centering
  \includegraphics[width=.95\linewidth]{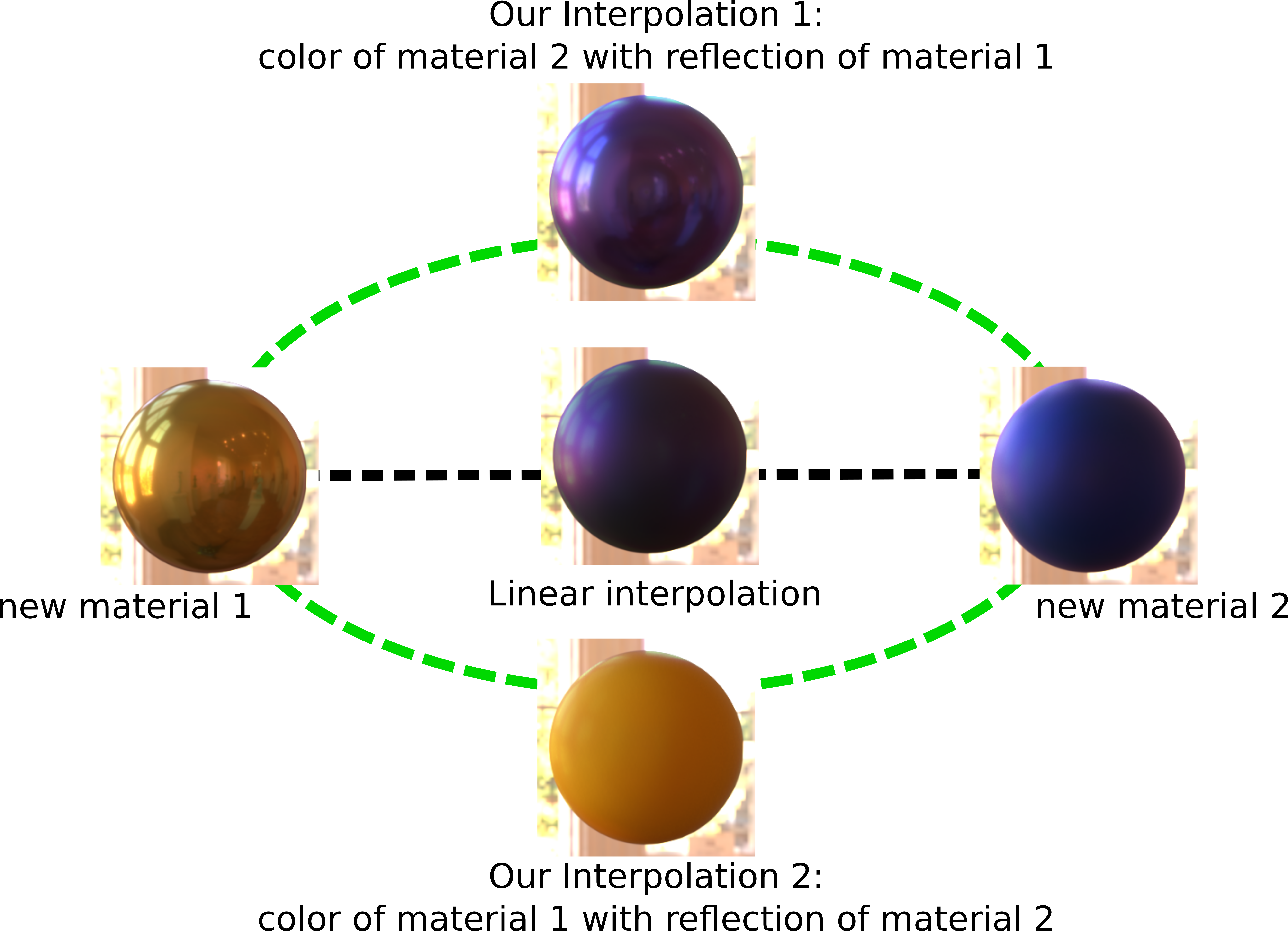}

  \caption{\label{fig:interp}
           A flexible interpolation between two very different materials. Our parametrization allows flexible interpolation between materials. Material 1 and material 2 are two new materials created with our parametrization and not originally present in the MERL dataset. For interpolation 1, we set the parameters controlling the color to the value of one material while setting the one representing the shape of the BRDF to the other material. We do the opposite for interpolation 2. For linear interpolation, all parameters are set to the mean value between parameters of material 1 and material 2.}
\end{figure}

\begin{figure*}[h]
  \centering
  \mbox{} \hfill
\begin{center}
\begin{tabular}{c c c c} 
 \includegraphics[width=.22\linewidth]{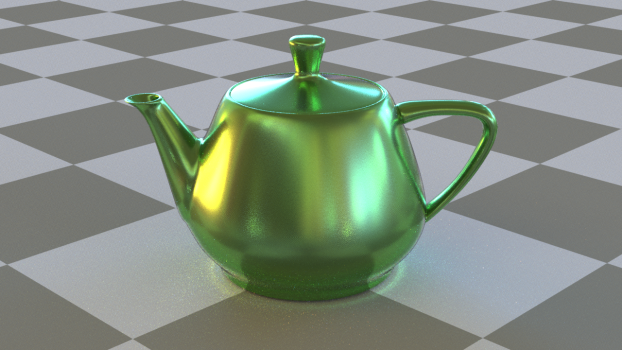} & 
 \includegraphics[width=.22\linewidth]{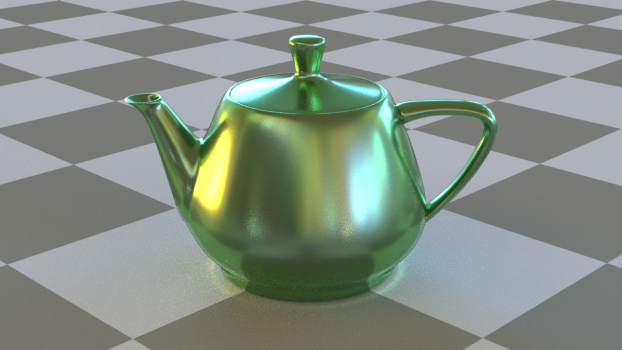}&
 \includegraphics[width=.22\linewidth]{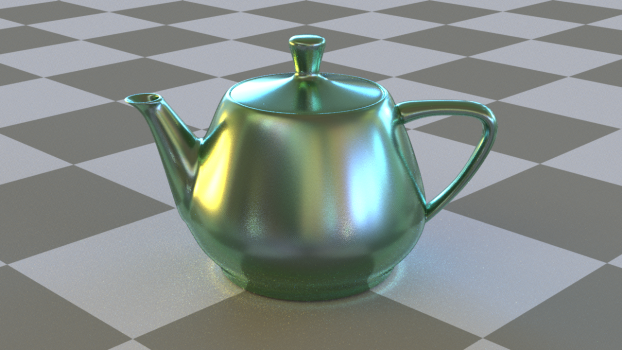}&
 \includegraphics[width=.22\linewidth]{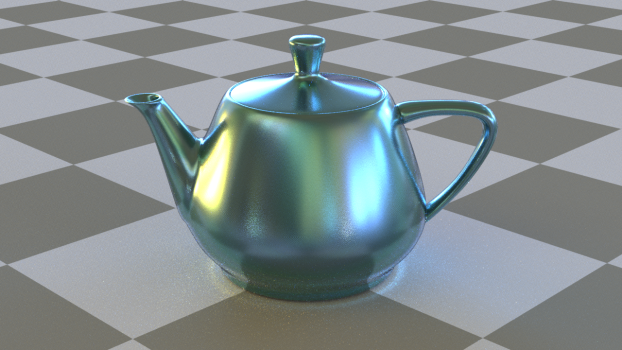} \\

 \includegraphics[width=.22\linewidth]{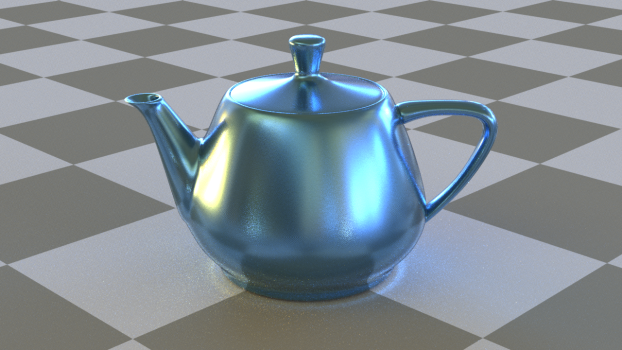} & 
 \includegraphics[width=.22\linewidth]{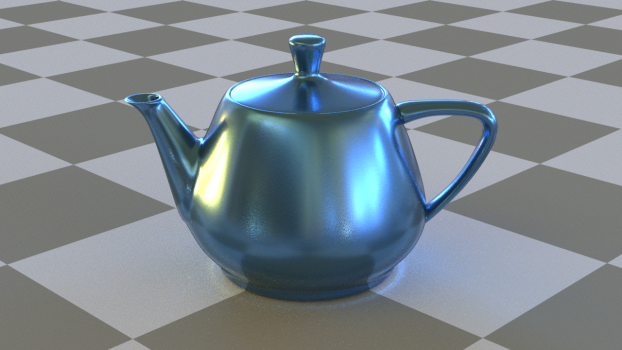}&
 \includegraphics[width=.22\linewidth]{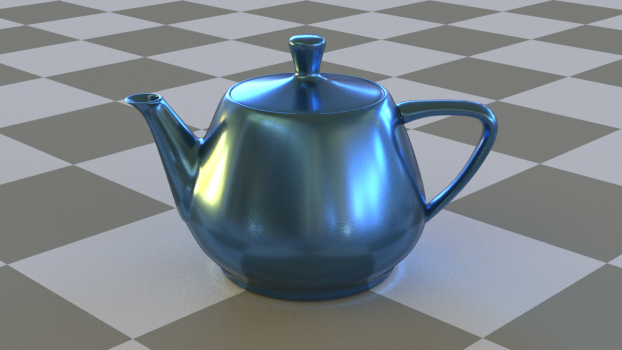}&
 \includegraphics[width=.22\linewidth]{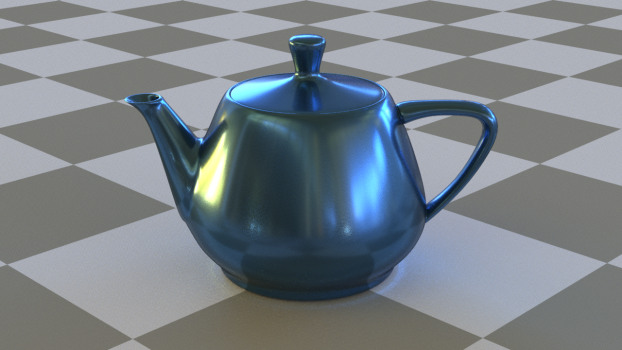} \\
 
   \includegraphics[width=.22\linewidth]{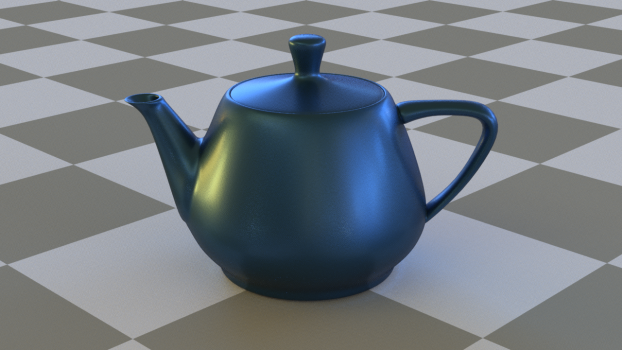} & 
 \includegraphics[width=.22\linewidth]{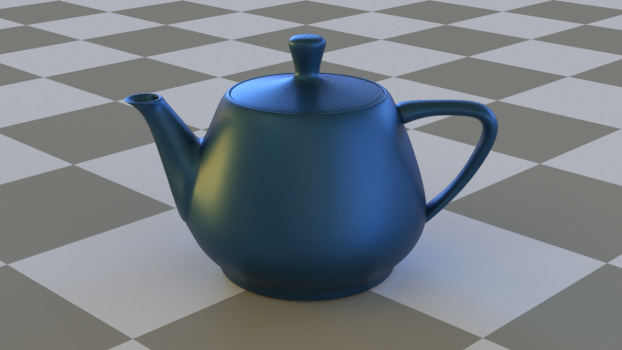}&
 \includegraphics[width=.22\linewidth]{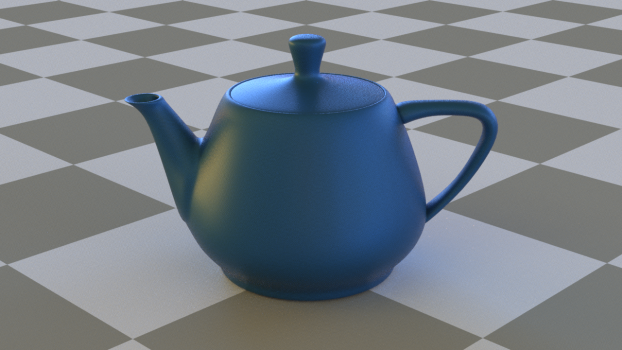}&
 \includegraphics[width=.22\linewidth]{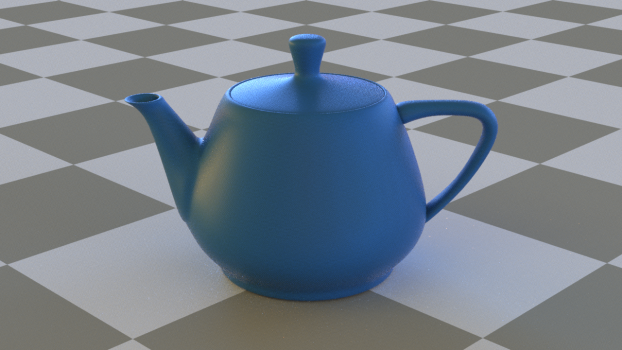} \\
 
  \includegraphics[width=.22\linewidth]{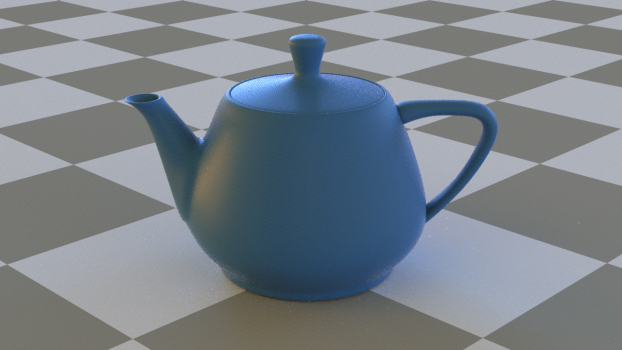} & 
 \includegraphics[width=.22\linewidth]{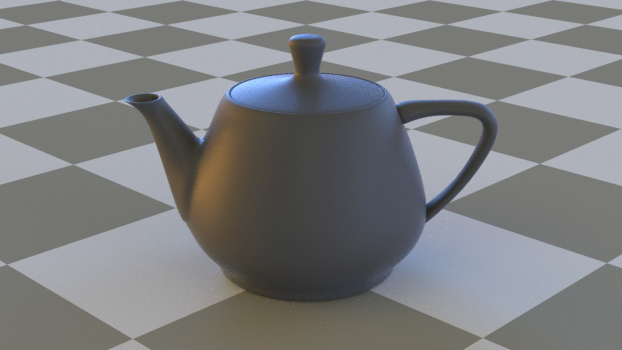}&
 \includegraphics[width=.22\linewidth]{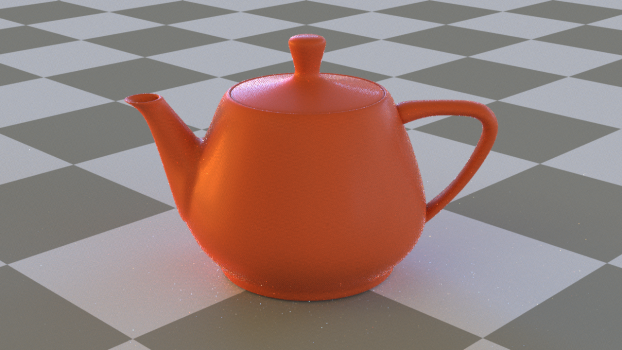}&
 \includegraphics[width=.22\linewidth]{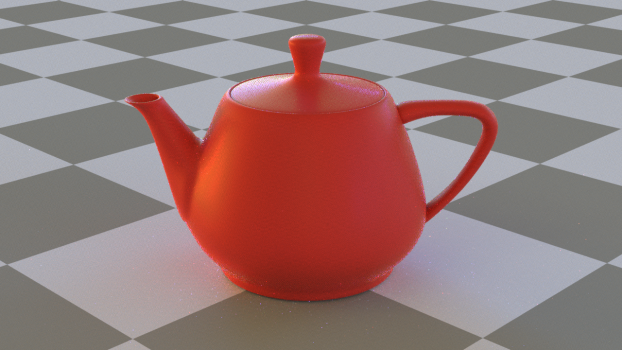} \\
 
\end{tabular}
\end{center}
\caption{\label{fig:easy_interpolation} Material editing from hazy green specular (top left) to red diffuse (bottom right).
First row we change the color from green to blue by modifying parameters $n^{o}$9 and $n^{o}$10 (decrease the green contribution) and parameters $n^{o}$1 and $n^{o}$8 (increase blue contribution). 
Second row we reduce the haziness by changing parameter $n^{o}$6. 
Third row we transition from specular to diffuse using parameter $n^{o}$5. Last row we modify the diffuse color from blue to red using parameters $n^{o}$1 and $n^{o}$9.}
\end{figure*}

\begin{figure*}[tbp]
  \centering
  \mbox{} \hfill
\begin{center}
\begin{tabular}{c c c} 
 \includegraphics[width=.3\linewidth]{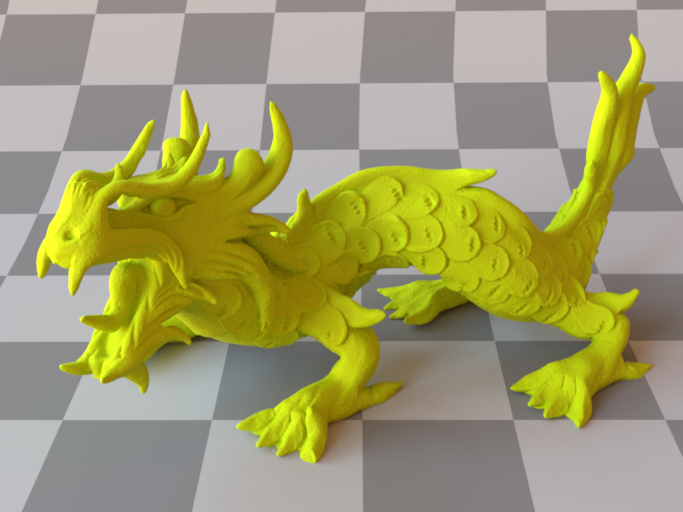} & 
 \includegraphics[width=.3\linewidth]{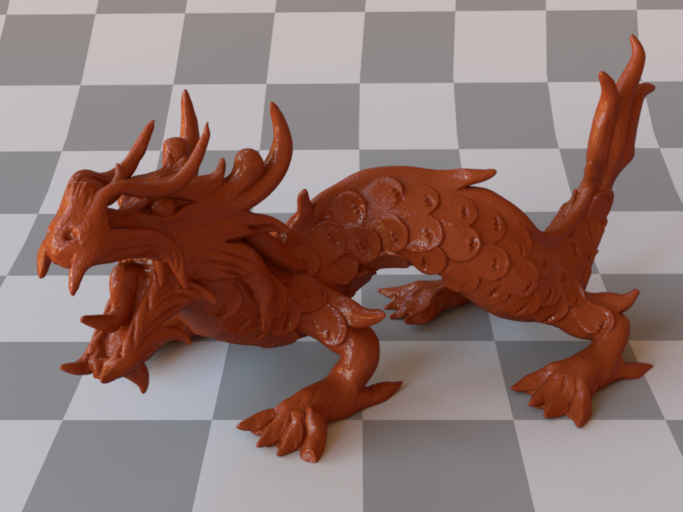}&
 \includegraphics[width=.3\linewidth]{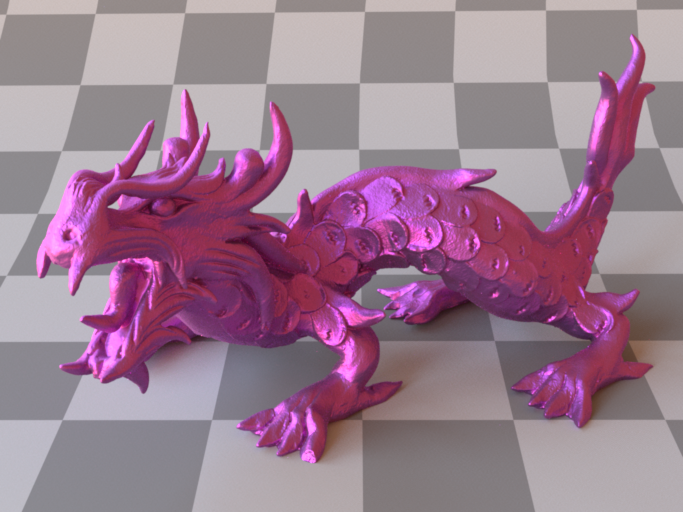}
\end{tabular}
\end{center}
\caption{\label{fig:new_mat} Creation of new material using our parametrization. Our interpretable parametrization allows the creation of new materials not present in the original dataset.}
\end{figure*}

In contrast, our new parametrization provides much more flexibility and precise control over the interpolation of the materials as illustrated Figure~\ref{fig:interp} and Figure~\ref{fig:easy_interpolation}. Instead of finding a path on a manifold for which we cannot precisely know which generative factor will be modified, our parametrization allows the precise control over specific generative factors. Figure~\ref{fig:easy_interpolation} and Figure~\ref{fig:teaser} illustrate the precise editing and rich rendering capabilities of our approach. 
Finally, our parametrization enables the creation of new materials not initially present in the dataset in a controlled fashion. Examples of new material can be seen Figure~\ref{fig:new_mat}.

We believe our interpretable disentangled parametrization is a sensible solution toward closing the gap between analytical and measured BRDF models. It offers both good control over editing distinct generative appearance factors and high variance in the possible producible appearances.



\section{Limitations and Future Work}
We would like to discuss several limitations of our current parametrization that we would like to solve in future work.
Also our set of parameters are interpretable, the practicability of the parametrization for artistic purposes is currently circumscribed. In future work, we would like to modify the training pipeline to make our parametrization more practical.
We have identified, two issues with the current disentanglement of our parameter space that require further investigations, the modularity and explicitness of the parametrization.
For the modularity, we identified leaks of some parameters through some others. For example we noticed that the color control is slightly leaking on to parameters $n^{o}$7 and $n^{o}$4. It can be observed in Figure~\ref{fig:latent_traversal}. In future work we would like to prevent those leaks by improving our latent space representation like in \cite{leaks1} or \cite{leaks2}.
The other limitation of our network is its explicitness. For example, our DNN cannot reproduce green by itself. The green reconstruction issue that our DNN suffers from initially was already present in \cite{DeepBRDF}. They reported a RelAE of $0.086$ for green-acrylic where we obtain $0.077$ Figure~\ref{fig:reconstruc_amelio}. However, thanks to the interpretability of our parametrization, and this is a genuine novelty of our work, we can significantly improve the explicitness by manually adding dimensions to the latent space. We thus reach a RelAE of $0.028$ with our final parametrization. Nonetheless, in future work, we would like to improve the natural explicitness of our network. Finally, this work is focused on the MERL dataset. Among many opportunities to consider for future developments, extending the approach to other material databases \cite{Dupuy2018Adaptive}, \cite{UTIA} would be an interesting option.


\section{Conclusion}
In an effort to combine the best of both worlds between the high variance of measured BRDF models and the high control parametrization of analytical BRDF models, we have developed a novel deep learning-based approach to create an interpretable disentangled parametrization space for measured BRDF. Our method relies on training a $\beta$-VAE and does not require a test-subject investigation. We have illustrated the visual interpretability of our parameters as well as the editing capabilities of our approach for reconstruction improvement or material editing. Furthermore, we have demonstrated that our set of learned parameters could be enriched with manual addition of parameters to extend the range of conceivable appearances. We believe that our approach leverages the great capacities of deep learning while removing one of its major drawback, the lack of interpretability. 

\bibliographystyle{eg-alpha-doi} 
\bibliography{egbibsample}       

\end{document}